\documentclass[twocolumn,preprintnumbers,amsmath,amssymb]{revtex4}
\usepackage{epsfig}
\def\be{\begin{equation}}
\def\ee{\end{equation}}
\def\o{\Omega_{b}}
\def\bq{\begin{eqnarray}}
\def\eq{\end{eqnarray}}
\def\rdr{\frac{\dot{a}}{a}}
\def\rdrsq{\frac{\dot{a^2}}{a^2}}

\def\l{\Lambda}

\begin{document}
\title{Stabilization of branes in a cosmological setting} 
\author{H.~K.~Jassal \thanks{E-Mail: hkj@iucaa.ernet.in}}
\affiliation{Inter University Centre for Astronomy and Astrophysics, \\
Post Bag 4, Ganeshkhind, Pune-411 007, India.}

\begin{abstract}
We  study  some cosmological consequences of the five
dimensional, two brane Randall-Sundrum scenario.
We integrate over the extra dimensions and in four dimensions the
action reduces to that of scalar tensor gravity.
The radius of the compact extra dimension is taken to be time
dependent.
It is shown that the radius of the extra dimension rapidly approaches
a constant nonzero separation  of branes.
A radion dominated universe cannot undergo accelerated expansion in
the absence of a potential.
It is shown that a simple quadratic potential with minimum at zero
leads to constant nonzero separation of branes in a similar level but
now accelerated expansion is possible.
After stabilization the quadratic
potential contributes an effective cosmological constant term.
We show that with a suitable tuning of parameters the requirements for
solving the hierarchy problem and getting an effective dark energy can
be satisfied simultaneously. 
%\pacs{ 04.50.+h, 95.30.SfA}
\end{abstract}

\maketitle

\section{introduction}
In this paper, we study some cosmological implications of
five-dimensional warped geometry  proposed by Randall and Sundrum 
\cite{randall1}. 
The Randall-Sundrum model consists of  two four dimensional 
branes which are defects in a five dimensional anti-deSitter
background.   
One of the branes is a positive tension Planck brane and the other is
the brane on which standard model particles are confined; this has a
negative tension and is called the TeV brane.  
The hierarchy between the four dimensional Planck scale and the
fundamental scale of the theory is resolved because of the presence of 
the exponential warp factor.

The five-dimensional spacetime is a slice of anti-deSitter geometry,
where we have a negative cosmological constant.
Two 3-branes  are located at fixed points of orbifold $S^{1}/Z_{2}$.
In other words, the extra fifth dimension is a circle with opposite
points identified.
We take the two orbifold points to be situated
at $y=0$ and $y=1/2$ \cite{csaki}; the positive tension Planck brane
is located at $y=0$ and the negative tension TeV brane is situated at
$y=1/2$.  

The action for the five dimensional anti-deSitter spacetime is given by 
\bq
\label{eq:fiveaction}
S &=& 2 \int d^4x \int_{0}^{1/2} dy \sqrt{-G} \left(M^{3} R-\l \right)
\\ \nonumber
&+&\int d^4x \sqrt{-g^{(+)}} \left(L^{+}-V^{+} \right)  \\ \nonumber
&+&\int d^4x \sqrt{-g^{(-)}} \left(L^{-}-V^{-} \right)
\eq
where 
\be
V^{+}=-V^{-}=12 m_{0} M^{3},~~\l=-12 m_{0}^{2} M^{3},
\ee
the five dimensional Ricci scalar is denoted by $R$, the bulk
cosmological constant is given by $\l$ and $M$ is the five 
dimensional Planck mass. 
The $(+)$ sign denotes the Planck brane and $(-)$ sign represents the
negative tension TeV brane.
The matter fields on the positive and negative tension branes are
$L^{+}$ and $L^{-}$ respectively, while $V^{\pm}$ represent the brane
tensions on positive and negative tension branes respectively.

The metrics on the two four-dimensional branes are therefore given by
\bq
g_{\mu \nu}^{(+)} = G_{\mu \nu}(x^{\mu}, y=0)~~{\rm and} \\ \nonumber
g_{\mu \nu}^{(-)} = G_{\mu \nu}(x^{\mu},y=1/2)
\eq
The five-dimensional Einstein equations are solved by the metric
\be
ds^2=e^{-2 m_0 r_c \mid y \mid} \eta_{\mu \nu} dx^{\mu} dx^{\nu} +
r_{c}^{2} dy^{2}
\ee
where $\eta_{\mu \nu}$ represents the flat four dimensional 3-brane
while $r_c$ is the radius of the extra dimension.
This model solves the mass hierarchy problem in particle physics and
has therefore been a subject of extensive study.

To study cosmology of the brane worlds, the radius of the extra
dimension is taken to be time dependent.
It was shown that in order that the brane world models be
consistent with observations, the separation between the branes, the
radion, should be a constant \cite{garriga}. 
In general,  the presence of bulk scalar fields achieves this constant
separation \cite{gw}.
In four dimensions the theory reduces to that of a
scalar tensor gravity with the Brans-Dicke factor which is a function
of the scalar field \cite{scalar} (for a review on brane
cosmologies, see \cite{brax}). 
In view of the present observations and belief that the universe is
undergoing an accelerated expansion (for a review see \cite{paddy}),
we investigate if the stabilizing potential provides the cosmological
constant contribution  to the energy density of the universe.
For various scalar field models of dark energy see
Refs. \cite{phi,tachyon,kessence,phantom,branedark}.

The paper is organized as follows. 
In Section II we study cosmology of the brane world model. 
It is shown that with a simple quadratic radion field  potential with
minimum at zero, we achieve stabilization as well as late time
acceleration in the evolution. 
Section III shows that a quadratic potential with nonzero minimum
changes the picture and we have the Hubble parameter oscillating about
the average evolution.
The main conclusions of this paper are summarized in Section IV.

%%%%%%%%%%%%%%%%%%%%%%%%%%%%%%%%%%%%%%%%%%%
\section{cosmology in brane world scenario}
%%%%%%%%%%%%%%%%%%%%%%%%%%%%%%%%%%%%%%%%%%%
For cosmological solutions, we assume the modulus $r_c$ to be time
dependent \cite{cosmo1,csaki,debchou}. 
The five-dimensional metric ansatz is \cite{csaki} 
\be
ds^2=e^{-2 m_{0} b(t) \mid y \mid} g_{\mu \nu} dx^{\mu} dx^{\nu} +
b^{2}(t) dy^{2}
\ee
with the four-dimensional spacetime being described by the spatially
flat Friedmann-Robertson-Walker metric
\be
g_{\mu \nu}={\rm diag} (-1, a^2(t), a^2(t), a^2(t))
\ee
where $a(t)$ is the scale factor.
This scale factor is different from what an observer on the negative
tension brane will see and  will be discussed later.

Using the above metric ansatz, we integrate over the extra dimension
in the action given in Eq. (\ref{eq:fiveaction}).
The four dimensional action can be written as
\bq
\label{eq:4dact}
S_{eff}&=&-\frac{3}{k^2 m_0} \int d^4 x \left[ \left(1-\o^2\right)
\rdrsq \right. \\ \nonumber
&+& \left.m_0 \o^2 \rdr \dot{b}-\frac{1}{4}m_0^2 \o^2 \dot{b^2}\right] \\
\nonumber
\eq
where $\o=e^{-m_0 b(t)/2}$ and $k^2=1/2M^3$.

In the action, we add a term $V(b(t))$, assuming a potential
associated with the radion field $b(t)$ which we will discuss later.    
The action can be further written as
\bq
\label{eq:4dact1}
S_{eff}&=&-\frac{1}{2 k^2 m_0}\int d^4x a^3(t) \left[(1-\o^{2} ) R_{4}
\right. \\ \nonumber
&-& \left.\frac{3}{2} m_{0}^2 \o^{2} \dot{b}^2 + V(b(t)) \right]
\eq
The four dimensional Ricci scalar is denoted by $R_4$. 
The action can be reduced to the standard Brans-Dicke scalar tensor
gravity if  we identify scalar field
$\phi=1-\o^2$, with Brans-Dicke factor $W(\phi)$ given by
$W(\phi)=\frac{3}{2}\frac{\phi}{1-\phi}$.

The cosmological equations of motion  obtained from this action are
\bq
3 \frac{\dot{a}^2}{a^2} &=&\frac{3 \dot{\o}^2}{1-\o^2} 
+6 \frac{\dot{a}}{a} \frac{\o \dot{\o}}{1-\o^2}
+\frac{1}{2} \frac{V(\o)}{1-\o^2} \\ \nonumber
2\frac{\ddot{a}}{a}+
\frac{\dot{a}^2}{a^2}&=&-\frac{\dot{\o}^2}{1-\o^2}+4 \frac{\dot{a}}{a} \frac{\o
\dot{\o}}{1-\o^2} 
+ \frac{2 \o \ddot{\o}}{1-\o^2} \\ \nonumber
&+&\frac{1}{2} \frac{V(\o)}{1-\o^2} \\ \nonumber
6 \frac{\ddot{\o}}{\o}&+&18\frac{\dot{a}}{a} \frac{\dot{\o}}{\o} = 2V+\frac{(1-\o^2)}
{2\o} \frac{dV}{d\o} 
\eq
where $H=\frac{\dot{a}(t)}{a(t)}$, $\o(t)=e^{-m_0 b(t)/2}$.

Only two of the above three equations are independent.
We transform variables to $\phi=1-\o^2$ and rewrite two of the 
equations
\bq
\label{eq:cosmoeq}
 \frac{\dot{a}^2}{a^2} &=& \frac{\phi}{4 (1-\phi)}
\frac{\dot{\phi}^2}{\phi^2} - \frac{\dot{a}}{a}
\frac{\dot{\phi}}{\phi} + \frac{1}{6} \frac{V(\phi)}{\phi} \\ \nonumber
\ddot{\phi} + 3 H \frac{\dot{\phi}}{\phi} &=& -\frac{1}{2}
\frac{1}{(1-\phi)} \dot{\phi^2} \\ \nonumber
&+&  (1-\phi)\frac{1}{3} \left[2 V(\phi) - \phi \frac{ d
  V(\phi)}{d \phi} \right]
\eq

\begin{figure}
\begin{center}
\epsfig{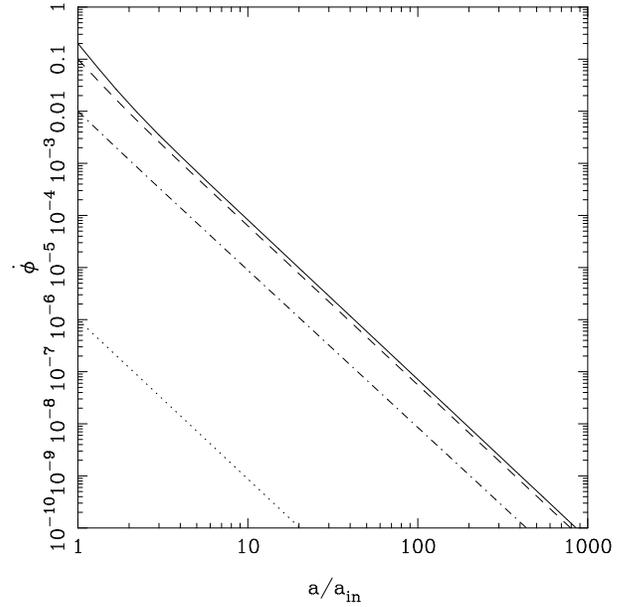}
\end{center}
\caption{The plot shows the variation of $\dot{\phi}$ with the scale
  factor.  It is clear in the figure that $\dot{\phi} \propto
  a(t)^{-3}$. This holds for no potential case  and as well as the
  case for a quadratic potential with a minimum at zero. The slight
  curve at very small $a/a_{in}$ is due to the $1-\phi$ term in
  Eq. \ref{eq:phidot}.} 
\label{fig:phiphase}
\end{figure}

The scalar field equation of motion in Eqs. \ref{eq:cosmoeq}  has a
critical point at $\phi=1$ and  $\dot{\phi}=0$ for a generic potential.
The branes can therefore stabilize only at infinite separation for
an arbitrary, featureless potential \footnote{Potentials with minima
  are discussed below}.
For a solution to the  hierarchy  problem, one needs the branes to be
at a finite separation.

In the absence of a potential, any value of $\phi$ is consistent
$\dot{\phi}=0$. 
Therefore the system converges to a finite separation of branes; the
final value of $\phi$ depends on the initial conditions.
For this case, the equation of motion for the scalar field can be
reduced to
\begin{equation}
\frac{d}{dt} \left[\frac{\dot{\phi}a^3}{\sqrt{1-\phi}}\right] = 0
\label{eq:phidot}
\end{equation}
It is clear from this expression that in most cases, the brane
separation stabilizes at the rate which inverse cube of the scale
factor. 
Therefore, the time taken for stabilization depends on the dominant
matter field in the universe.  
It consequently stabilizes hierarchy between the different scales (for
an extensive literature on modulus stabilization and the cosmological
implications see  \cite{gw,stable} and references therein).   
The derivative of the scalar field $\dot{\phi}$ drops to very small
value within a few expansion factors in the scale factor and therefore
from cosmological point of view may be considered to be negligible.
As $\phi$ approaches zero, the radion field stops contributing to the
cosmological evolution.
Even though $\dot{\phi} \neq 0$, and hence strictly
speaking the system does not reach a stable point and tends towards it
only asymptotically, however, as the rate of change of the brane
configuration becomes very small compared to the hubble parameter in a
very short time, the system is essentially not changing. 
This is indeed a remarkable result in that the brane system is
stabilized to some value by the cosmological expansion, but the radion
field does not affect the cosmological expansion after reaching a
stable configuration. 

\section{quadratic potential}

\subsection{Potential with minimum at zero}
From the scalar field equation we conclude  that if we have a
quadratic potential, $V(\phi)=V_o \phi^2$  (see also \cite{cline}), 
the term in square brackets in Eqs. \ref{eq:cosmoeq} vanishes and the
branes stabilize at any separation.
We get the same evolution behaviour for $\dot{\phi}$ as in
Eq. \ref{eq:phidot}. 
However, unlike the no potential case, the quadratic potential
contributes a cosmological constant term in the Friedmann equations. 
This is a desired feature as the present observations indicate that
the expansion of the universe is accelerating \cite{perl}.
Thus the quadratic potential can solve two problems at the same time.

Once the radion field settles down,  the scale factor and time $t$ are
scaled by a constant. 
The scale factor obtained in these coordinates after stabilization has the
same interpretation as the scale factor $Y(\tau) = e^{-m_0b(t)/2}
a(t)$  from the point of the view of an observer on the brane.
The equations of motion then remain the same with the factor $V_0$
scaled by $e^{m_0 b_0}$ where $b_0$ is the value of the stabilized
field. 
We can therefore consider the scale factor $a(t)$ to be describing
the `physical' scale factor.

\subsection{Numerical solutions with a quadratic potential}
We solve the equations of motion with potential $V = V_0 \phi^2$
numerically to substantiate the points discussed above.  
We rescale the equations with the initial  Hubble parameter by making
the following change $x = t H_{in}$ and $y= a/a_{in}$. 
We consider the presence of nonrelativistic and relativistic matter
only on the four dimensional negative tension brane. 
The equations transform to
\bq
\frac{y'^2}{y^2} &=& \frac{\Omega_{M_{in}}}{y^3} +
\frac{\Omega_{R_{in}}}{y^4} + \frac{\phi}{4 (1-\phi)}
\frac{\phi'^2}{\phi^2} \\ \nonumber 
&-& \frac{y'}{y}
\frac{\phi'}{\phi} + \frac{1}{6} \frac{V(\phi)}{\phi H_{in}^2} \\ \nonumber
\ddot{\phi} + 3 \frac{y'}{y} \frac{\phi'}{\phi} &=& -\frac{1}{2}
\frac{1}{(1-\phi)} \phi'^2 \\ \nonumber 
&+&  (1-\phi)\frac{1}{3 H_{in}^2} \left[2 V(\phi) - \phi \frac{d
    V(\phi)}{d \phi} \right] 
\eq
where $\Omega_{M_{in}}, \Omega_{R_{in}}$ are density parameters
for nonrelativistic and  relativistic matter respectively.
The initial value of $x$ is arbitrary, $y_{in}=\frac{a}{a_{in}}=1$
and we vary the initial values of $\phi$ and $\phi'$ for a given value
of $\Omega_\phi$. 
As mentioned above, we consider the quadratic potential $V=V_0\phi^2$.
From the structure of equations, it is clear that the amplitude $V_0$
always appears in the combination $\alpha=V_0/H_{in}^2$ and this is
how we choose to parameterize its value.
We consider the quadratic potential mentioned above and fix the
parameter $\alpha=V_0/H_{in}^2$ by  
\bq
\alpha=\frac{6}{\phi} \left[ \Omega_{\phi_{in}} -
  \frac{\phi'^2}{4 (1- \phi) \phi^2} + \frac{\phi'}{\phi}\right]
\eq

\begin{figure}
\begin{center}
\epsfig{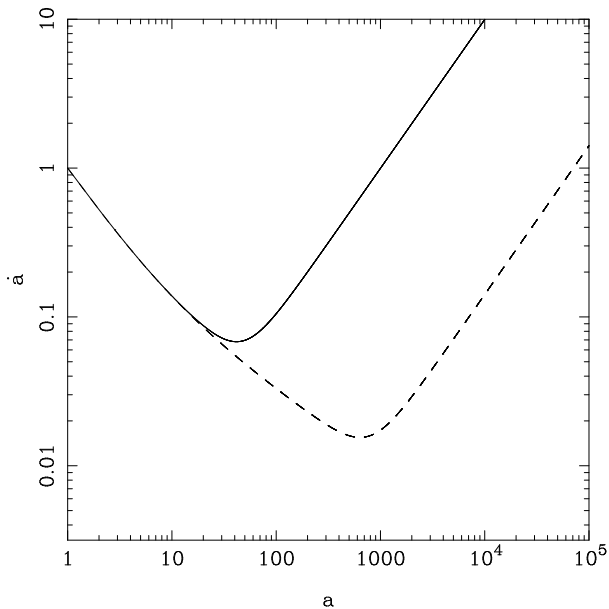}
\end{center}
\caption{Phase plot for the scale factor. Dashed lines are for
  $\alpha_{in}= 10^{-10}$,   $\dot{\phi}=10^{-10}$ and
  $\Omega_{M_{in}}=0.0001$. Solid lines   are for $\alpha_{in}=
  10^{-6}$,   $\dot{\phi}=10^{-6}$, $\phi = 1 -   e^{-10}$ and
  $\Omega_{M_{in}}=0.0001$.}   
\label{fig:newphase}
\end{figure}

Fig. \ref{fig:phiphase}  shows the behaviour of $\dot{\phi}$ as a
function of the scale factor and it is shown that $\dot{\phi} \propto
a(t)^{-3}$. 
Here we are dealing with a stable attractor at $\dot\phi=0$ for the
quadratic potential.
The scalar field stabilizes to its asymptotic value within a few
expansions factors and it behaves like cosmological constant this
point onwards.
The effective cosmological constant is proportional to $\alpha$,
therefore accelerating phase sets in early if the value of this
parameter is large. 
After this the rate of expansion continues to accelerate unless one
introduces additional coupling with matter to the field. 
If this parameter is small, then there is late time acceleration and
the potential provides the dark energy component to the energy density
of the universe. 
The value of the cosmological constant depends on the initial
conditions. 
Therefore the initial conditions need to be tuned in order
to get the desired solutions.
This point becomes clear in  Fig. \ref{fig:newphase}  where we have
shown the phase plot for the scale factor.

%%%%%%%%%%%%%%%%%%%%%%%%%%%%%%%%%%%
\subsection{Coordinate transformation}
%%%%%%%%%%%%%%%%%%%%%%%%%%%%%%%%%%%
For a cosmological interpretation of the scale factor from the point
of view of an observer on the brane, we transform the variables as 
 (if there is no potential and the scalar field phi
is the dominant constituent of the universe) \cite{debchou}
\be
d\tau^2= e^{-m_0b(t)} dt^2;~~~~~Y(\tau)=e^{-m_0b(t)/2} a(t)
\ee
Again for a quadratic potential and for no potential case, the scalar
field equation of motion leads us to  
$$
\frac{d}{dt}\left[\frac{\varphi' a^3} {\sqrt{1 + \alpha
      \varphi}} \right]=0, 
$$
where $\varphi= \frac{1-\o^2}{\alpha \o^2}$  (see for details, \cite{debchou}), 
$\alpha$ being a constant.
The field shows the same behaviour with respect to the scale factor as the one
in the coordinate system considered earlier. 
Here, prime represents derivative with respect to the variable
$\tau$.

%%%%%%%%
% No potential
%%%%%%%%
The equation for the Hubble parameter changes to (if there is no potential)
\be
\frac{Y'^2}{Y^2} = - \frac{Y'}{Y} \frac{\phi'}{\phi}-\frac{\alpha
  \phi}{4(1+\alpha \phi)} \frac{\phi'^2}{\phi^2}
\ee
This equation implies that for an expanding universe, we need a
negative $\dot{\phi}$.
The second Friedmann equation can be recast in the form 
\be
2\frac{Y''}{Y} = \frac{Y'}{Y} \left(\frac{\varphi'}{\varphi} - \frac{Y'}{Y}
\right) 
\ee
For the universe to accelerate, we need  $Y''/Y > 0$.
This means that $\varphi'>0$ should also be satisfied for accelerated
expansion along with the above condition, and the Hubble parameter must
satisfy the constraint $H<\frac{\varphi'}{\varphi}$.
An expanding universe and a positive $\phi'$ are not compatible with each
other.
Therefore, one cannot get
inflationary solutions driven by the brane system in the absence of a
potential.
Of course these restrictions can be circumvented by invoking a potential.

\subsection{Potential with a nonzero minimum} 
If we wish to stabilise the branes at a particular finite separation,
we can use a potential with a minimum at the relevant value of
$\phi$. 
This replaces the need for special initial conditions with a tailored
potential to guide the brane system towards the desired asymptotic state. 
We consider a quadratic potential with a minimum at $\phi_0)^2$. 
The equations of motion then take the form
\bq
\frac{\dot{a}^2}{a^2}+\frac{\dot{a}}{a} \frac{\dot{\phi}}{\phi} &=&
\frac{\dot{\phi}^2}{4(1-\phi)\phi} + \frac{V_0 (\phi-\phi_0)^2}{6
  \phi} \\ \nonumber
\ddot{\phi}+ 3 \frac{\dot{a}}{a} \dot{\phi} &=& -\frac{1}{2}
\frac{\dot{\phi}^2}{(1-\phi)} \\ \nonumber &-& \frac{2}{3}(1-\phi) V_0 \phi_0
(\phi-\phi_0) 
\eq
The scalar field rolls down the potential and undergoes oscillations
near the minimum of the potential. 
To look at the behaviour of Hubble parameter around this point we make
the  perturbation expansion
\bq
H=\bar{H}+\delta,~~~~\phi=\phi_0+\epsilon
\eq
which up to the first order in $\delta$ and $\epsilon$ reduce to
\bq
\bar{H}^2 + 2 \delta \bar{H} + \bar{H}
\frac{\dot{\epsilon}}{\phi_0} &=& \frac{8 \pi G}{3} \rho \\ \nonumber
\ddot{\epsilon} + 3 \bar{H} \dot{\epsilon} &=& \frac{2}{3} V_0 \phi_0
(1-\phi_0) \epsilon
\eq
where $\rho$ is the energy density of the matter fields.
The zeroth order equation describes the evolution of the non-oscillatory
component of the Hubble parameter $\bar{H}$.
This component is
insensitive to the scalar field $\phi$ and its evolution is governed by
the radiation/matter fields on the brane.  The first order equations
describe the coupled behaviour of the field $\phi$ and the oscillatory
component of the Hubble parameter.  The scalar field undergoes damped
oscillations about the minimum of the potential with $3\bar{H}$ being
the damping coefficient.  The oscillatory component of Hubble
parameter $\delta$ is proportional to the velocity of the field
$\phi$ and is given by $\delta = - \dot{\epsilon}/2\phi_0$.    
These oscillations are undesirable from a
cosmological point of view and this toy model is therefore, less than
satisfactory.  However, adding a coupling between $\phi$ and matter fields may
lead to damping of oscillations in the Hubble parameter.

\section{summary}

This paper presents some cosmological solutions allowed by the five
dimensional Randall-Sundrum two brane scenario.
We show that the branes are stabilised to a finite seperation.
A stable point for brane separation is  of interest if this
point is reached sufficiently rapidly.  
The separation of branes is tied up with fundamental constants in
particle physics and there is little evidence to show that these have
changed much between the present epoch and very high redshifts ($z
\approx 10^{10}$). 
Thus the separation between branes must reach its stable value at
sufficiently high redshifts, long before the effective cosmological
constant that it provides becomes the dominant constituent of the
universe in terms of energy density.
We summarise the main results for the cosmology of the model discussed
above 
\begin{itemize}
\item The stable configuration is achieved fairly quickly, i.e. the
derivative of the radion field vanishes rapidly. 
\item After the stabilization is achieved there is no further
  cosmological consequence of the radion field. 
\item If we  assume a simple quadratic potential for the
radion field,  radius stabilization can be achieved and the scalar
field evolution equation has the same form as in no potential case.
\item In addition to this stabilization, the potential also provides the dark
energy component at late times.
\item 
If we want to solve the hierarchy problem as well as the dark energy
problem using a quadratic potential then fine tuning of initial
conditions is required.  
The level of fine tuning required is similar to that in other
models of dark energy. 
\end{itemize}
The small rate of change of $\dot{\phi}$  may reflect itself in
changing fundamental constants but clearly the rate of change will be
much smaller than the expansion rate of the universe. 
The fact that we achieve stabilization and dark energy component from
the same potential is an attractive feature and hence this model may
be a candidate for further study.

\section*{Acknowledgements}
The author thanks J. S. Bagla, T. Padmanabhan and  K. Subramanian for useful
discussion.  
Thanks are also due to  D. Choudhury and D.
Jatkar for discussion on brane worlds at various stages.

\end{document}